\documentclass[aps,superscriptaddress,twocolumn]{revtex4}

\usepackage{graphicx,graphics,epsfig}   
\usepackage{dcolumn}    
\usepackage{bm}         
\usepackage{amsmath}    
\usepackage{verbatim}   
\usepackage{color}      
\usepackage{times,natbib}
\usepackage{amsmath,amsfonts,amssymb,graphics,graphics,color,times}
\usepackage{bm}

\usepackage{latexsym}
\usepackage{amsmath}
\usepackage{amssymb}
\usepackage{amsfonts}
\usepackage{amsthm}
\usepackage{mathrsfs}
\usepackage{color,verbatim,graphics}
\DeclareMathAlphabet{\mathrsfs}{U}{rsfs}{m}{n}
\DeclareMathAlphabet{\mathpzc}{OT1}{pzc}{m}{it}
\DeclareMathAlphabet{\matheus}{U}{eus}{m}{n}
\DeclareMathAlphabet{\mathbbold}{U}{bbold}{m}{n}

\def\r1{\textbf{r}}

\def\nr{{\bf n}_R^{}}

\newcommand{\ba}{\begin{eqnarray}}
\newcommand{\ea}{\end{eqnarray}}
\newcommand{\ban}{\begin{eqnarray*}}
\newcommand{\ean}{\end{eqnarray*}}
\newcommand{\be}{\begin{equation}}
\newcommand{\ee}{\end{equation}}

\newcommand{\ket}[1]{|#1\rangle}

\begin{document}

\title{Polarization engineering in photonic crystal waveguides for spin-photon entanglers }

\author{A. B. Young}
\affiliation{Department of Electrical and Electronic Engineering, University of Bristol,
Merchant Venturers Building, Woodland Road, Bristol, BS8 1UB, UK}
\author{A. C. T. Thijssen}
\affiliation{Centre for Quantum Photonics, H.H. Wills Physics Laboratory, University of Bristol,
Tyndall Avenue, Bristol, BS8 1TL, United Kingdom}
\author{D. M. Beggs}
\affiliation{Centre for Quantum Photonics, H.H. Wills Physics Laboratory, University of Bristol,
Tyndall Avenue, Bristol, BS8 1TL, United Kingdom}
\author{P. Androvitsaneas}
\affiliation{Centre for Quantum Photonics, H.H. Wills Physics Laboratory, University of Bristol,
Tyndall Avenue, Bristol, BS8 1TL, United Kingdom}
\author{L. Kuipers}
\affiliation{Center for Nanophotonics, FOM Institute AMOLF, Science Park 104, 1098 XG Amsterdam, The Netherlands.}
\author{J. G. Rarity}
\affiliation{Department of
Electrical and Electronic Engineering, University of Bristol,
Merchant Venturers Building, Woodland Road, Bristol, BS8 1UB, UK}
\author{S. Hughes}
\affiliation{Department of Physics, Queen's University, Ontario, Canada K7L 3N6}
\author{R. Oulton}
\affiliation{Centre for Quantum Photonics, H.H. Wills Physics Laboratory, University of Bristol,
Tyndall Avenue, Bristol, BS8 1TL, United Kingdom}
\affiliation{Department of Electrical and Electronic Engineering, University of Bristol,
Merchant Venturers Building, Woodland Road, Bristol, BS8 1UB, UK}

\begin{abstract}
By performing a full analysis of the projected  local density of states (LDOS) in a photonic crystal waveguide, we show that phase plays a crucial role in the symmetry of the light-matter interaction.  By considering a quantum dot (QD) spin coupled to a photonic crystal waveguide (PCW) mode, we demonstrate that the light-matter interaction can be asymmetric, leading to unidirectional emission and a deterministic entangled photon source. Further we show that understanding the phase associated with both the LDOS and the QD spin is essential for a range of devices that that can be realised with a QD in a PCW. We also show how quantum entanglement can completely reverse photon propagation direction, and highlight a fundamental breakdown of the semiclassical dipole approximation for describing light-matter interactions in these spin dependent systems.

\end{abstract}

\maketitle

Nanophotonic structures are routinely used to enhance light-matter interactions by modifying the density of electromagnetic (EM) field modes. This is often simplified to a scalar quantity, the local density of states (LDOS). However we show that the EM field modes also contain important phase information,
which interacts with a  phase-dependent emitter in a non-trivial, non-intuitive way. This extra phase information is vital in practical designs of integrated quantum photonic circuits, a leading contender for future 
quantum technologies~\cite{Politi:2008fk}.  

In a quantum photonic circuit, information may be stored and transmitted via photons, which make excellent flying qubits. Photons suffer little from decoherence, and single qubit gates performed by changing photon phase are straightforward.  Less straightforward is the ability to create two qubit gates, where one photon switches another's state: direct photon-photon interactions are extremely weak. One type of matter system which has potential to mediate photon-photon interactions is a quantum dot (QD) which acts as an artificial atom. Its solid-state nature means that it is relatively simple to enhance the light-matter interaction by incorporating it into microcavity structures. Simultaneously, a sizeable research effort into using the electron spin state in QDs has shown much success.  In particular, the long spin coherence times ($\mu$s)~\cite{Greilich:2006uq,De-Greve:2011ys}, and ease of optical initialisation, coherent control and readout have all been demonstrated~\cite{Carter:2013uq,De-Greve:2011ys,Press:2010vn}.  Thus the potential exists to use the QD spin as a static qubit in order to mediate deterministic photon-photon interactions.

If future devices are to be part of an  integrated quantum photonic chip then a promising platform is photonic crystal waveguides (PCW) and cavities~\cite{OBrien:2009kx}. A QD embedded in a PCW has already been recognised as an excellent single photon source~\cite{Manga-Rao:2007ly,Lecamp:2007ve,Yao:2010dq}, where highly efficient coupling between a QD exciton transition and the PCW has been demonstrated~\cite{Lund-Hansen:2008zr}. This is because PCWs are approximately ``one dimensional'', where most of the energy from the emitter couples to the waveguide. The natural consequence of this is that simple  ``one dimensional atom" models\cite{waks:153601,auffeves-garnier:053823} may be applied to a PCW. In this Letter, we consider the coupling between polarised spin-dependent transitions of a QD trion to a PCW.  We demonstrate that there is a complex interplay between the polarization structure of the PCWÕ mode, the QD spatial location and its spin state, leading to different functionalities that are not predicted by a schematic one-dimensional atom model.  This leads to surprising results, with different QD spatial locations enabling different quantum devices in the same waveguide.

\begin{figure}
\centering
\includegraphics[width=0.48\textwidth]{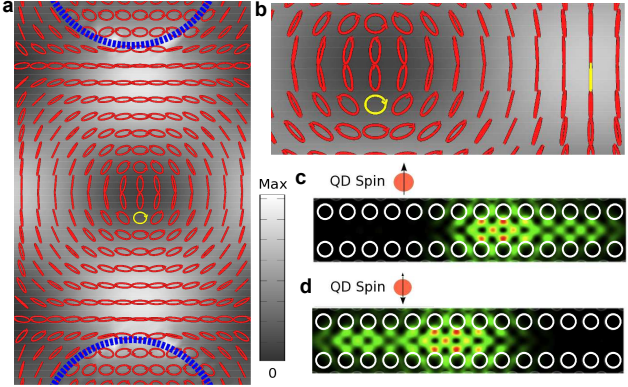}
\caption{(Color online) ({\bf a}) Zoom in of a W1 PCW made from a suspended slab of GaAs with air holes (marked with dashed blue line) lattice constant $a=250\,$nm and the hole size is $0.34a$. A line of holes is missing through the centre forming the waveguide. Grayscale background shows field intensity, red markings show polarisation ellipse, where straight lines represent linear polarisation. ({\bf b}) Zoom of specific area where the yellow circle represents the C-point and yellow line the ${\bf E}_y$ polarised point we consider in this paper. FDTD simulations showing emission from a negatively charged QD at the identified C-point for ({\bf c})  spin  up ($\sigma_+$ polarised), and ({\bf d})  spin down ($\sigma_-$ polarised)}\label{fig:FDTD}
\end{figure}

A two dimensional PC is formed from a slab of dielectric containing periodically spaced air-holes which modulate the refractive index, giving rise to a photonic bandgap. In plane confinement is provided by the photonic bandgap, which dramatically reduces the local density of states (LDOS) of optical modes, relative to bulk material, into which a dipole can emit~\cite{Lecamp:2007ve}. If a line defect consisting of a line of missing holes is incorporated, a waveguide is formed (see Fig.~\ref{fig:FDTD}a.). The propagation of light along the waveguide supports slow light modes~\cite{Baba:2008kx}, which increase the LDOS in the waveguide region. As a result, the dominant modes for dipole emission are into this region thus forming a one-dimensional ``wire-like'' waveguide structure \cite{Kleppner:1981tg}. In contrast, in a standard waveguide the bulk LDOS is not significantly modified, and light scattered from the emitter is mainly into leaky modes.

Another significant difference between a standard planar waveguide  and a PCW is the polarization state of the light propagating inside the structure. A standard waveguide supports a TE mode which is constant along the length of the guide. However, the PCW supports bound Bloch modes with significant components of both $E_x$ and $E_y$ fields, that vary strongly across one lattice period.  Hence different locations inside the PCW support different superpositions of ${ E}_x$, and ${ E}_y$ with a fixed relative phase that varies spatially. At each point the field may be expressed as a polarization ellipse, as shown in Fig.~\ref{fig:FDTD}(a). There are clearly points where the ellipse becomes circular which corresponds to a ``C-point'' singularity~\cite{Burresi:2009fk}, and also where the ellipse collapses to a line (L-line) where  the polarisation is linear. These are known collectively as polarization singularities~\cite{Nye:1983kx}. It is clear that the polarization of the mode is intricate, with an arbitrary point in the PCW (${\bf r}_0$) showing an arbitrary local electric field polarization, with ${\bf e}_k({\bf r}_0)= \alpha {\bf E}_x + e^{i\phi}\beta {\bf E}_y$.

The QDs themselves are modelled as point-like emitters. In addition,  negatively doped QDs with a resident electron spin undergo strict selection rules that couple to $\sigma_+$ circularly polarized light for spin up and $\sigma_-$ light for spin down. The QD emitter is dipole-like and thus these spin transitions may be modelled as superpositions of orthogonal dipoles aligned along $x$ and $y$, i.e. $\bm\mu =\alpha\bm\mu_x+e^{i\phi}\beta\bm\mu_y$, where $\bm\mu$ represents a unit vector in the dipole direction. In bulk or simple dielectric structures, the coupling strength of the emitter is calculated to be proportional to the scalar product of $|\bm\mu \cdot {\bf E}({\bf r}_0)|/ |\bm\mu|.|{\bf E}_{\rm max}|$, with the available local density of states (LDOS) proportional to $|{\bf E}_{\rm max}|^2$.  However, the LDOS does not contain the full phase information present in the EM field modes.  This necessitates a departure from this simple model and use of a Green function analysis~\cite{Hughes:2004oq, Wubs:2004bh,Yao:2010dq}, where the radiative coupling between the dipole and the waveguide mode is proportional to  $\bm\mu^{\dagger}\cdot {\bf G}(\bf r_0,\bf r_0)\cdot \bm\mu$, as outlined below.
The Green's function describes the
response at ${\bf r}$ to an oscillating dipole
at ${\bf r}_0$.

In the frequency domian,
the Green's function for the waveguide mode
is described through~\cite{Yao:2010dq} ($\omega$ is implicit)
 \begin{eqnarray}\label{gf}
 {\bf G}_{\rm w} ({\bf r}, {\bf r}_0)= {\bf G}_{\rm f}({\bf r}, {\bf r}_0)
 +{\bf G}_{\rm b}({\bf r}, {\bf r}_0)=\\\nonumber
    \frac{ia\omega}{2v_g }\left [\Theta(x-x_0){\bf e}_{k}({\bf r}){\bf e}_{k}^{*}({\bf r}_0)e^{ik(x-x_0)}
     +\right .\ \label{eq:Chap_5_DGF} \\\nonumber
        \left . \Theta(x_0-x){\bf e}_{k}^{*}({\bf r}){\bf e}_{k}({\bf r}_0)e^{-ik(x-x_0)}\right]
\end{eqnarray}

\noindent where $a$ is the lattice constant, $v_g $ is the group velocity, $\Theta$ is the Heaviside step function, $x_0$ is the $x$ coordinate of the dipole,  ${\bf e}_{k}(\bf r)$ is the propagating mode for wavenumber $k, $ normalized according to $\int_{V_c} \epsilon(\mathbf{r})|\mathbf{e}_{ k}(\mathbf{r})|^2d\mathbf{r}=1$, where $V_c$ is the spatial volume of a PC unit-cell, with $\epsilon({\bf r})$ the dielectric function. The first (second) term in Eq.~(\ref{gf}) represents the Green's function for the forwards (backward) propagating mode.
An arbitrary point in the PCW (${\bf r}_0$) will thus have a local electric field polarization ${\bf e}_k({\bf r}_0) = \alpha {\bf  E}_{x}+e^{i\phi}\beta {\bf E}_{y}$, for light that is propagating in a forwards propagating Bloch mode. Whereas in the backwards propagating Bloch mode,  ${\bf e}_k({\bf r}_0) = \alpha {\bf E}_{x}+e^{-i\phi}\beta {\bf E}_{y}$. We now consider a specific point in the PCW where the field is circular (C-point), i.e. where $\alpha=\beta$, and $\phi=\pi/2$. Here we find if one sets $\bm\mu=\bm\sigma_{+}$ then (excluding constants) $\bm\mu^\dagger \cdot {\bf G}_{\rm f}({\bf r}_0, {\bf r}_0)\cdot\bm\mu=1$ and  $\bm\mu^\dagger \cdot {\bf G}_{\rm b}({\bf r}_0, {\bf r}_0)\cdot \bm\mu=0$. Hence a right circularly polarised dipole will only couple to the forwards propagating mode. Similarly a left circularly polarised dipole will only couple to the backwards  mode.

The result is that at the C-point, there is a one-to-one correspondence between spin orientation and emission direction. To confirm this we perform in-house FDTD simulations of a W1 waveguide with slab thickness of $0.56a$ and hole radius of $0.34a$, where $ka/2\pi=0.39$ and $v_g=c/88$. In Fig.~\ref{fig:FDTD}c we consider an $\ket{\!\uparrow}$ ($\ket{\sigma_+}$ circular dipole) located at the C-point and in Fig.\ref{fig:FDTD}d. the spin is oriented $\ket{\!\downarrow}$ ($\ket{\sigma_-}$ circular dipole). Both show a unidirectional emission, dependent on spin orientation, in concurrence with the analytical Green function analysis above demonstrating 100\% unidirectionality. This striking result is due to the spin helicity in this system breaking the symmetry and allowing unidirectional emission. Recent work has shown partial spin path correlations in 
other structures~\cite{Rodriguez-Fortuno:2013fk,Luxmoore:2013ys}. We show here, for the first time to our knowledge, how to precisely engineer these correlations, which is in excellent agreement with recent measurements using near field microscopy techniques~\cite{Kobus_arxiv}.          
Spin-path entanglement is a natural consequence of this analysis. An  $\ket{\!\uparrow}$ dipole emits photons in the forward direction in the state $\ket{\rm f}$, while a $\ket{\!\downarrow}$ dipole emits photons in the backwards direction in  state $\ket{\rm b}$. An equal superposition of $\ket{\!\uparrow}+\ket{\!\downarrow}$ results in the output state:
\begin{equation}\label{cpem}  
\ket{\psi}_{\rm out}=\ket{\!\uparrow}\ket{\rm f}+\ket{\!\downarrow}\ket{\rm b},
\end{equation}
  an entangled state of photon path and spin orientation.

 The efficiency of the source is given by the $\beta$-factor, defined as
$\beta = \frac{\Gamma_{\rm w}}{\Gamma_{\rm w}+\Gamma_{\rm 0}}$,
where
$\Gamma_0$ represents radiative losses to modes above the light line; typically this latter contribution is much smaller than radiative decay to the waveguide mode, and is computed to be around $0.1\Gamma^{\rm hom}$, where $\Gamma^{\rm hom}$ represents the decay in the homogenous bulk material.
The coupling rate to waveguide modes,
$\Gamma_{\rm w}$, 
depends on the coupling to the projected LDOS. The rate of emission can be split into two parts: the rate forwards is given by $\Gamma_{\rm w}^{\rm f}=2d_0^2{\bm\mu}^\dagger\cdot {\bf G}_{\rm f}({\bf r}_0,{\bf r}_0)\cdot \bm\mu/\hbar\epsilon_0$ and the rate 
backwards,  
$\Gamma_{\rm w}^{\rm b}=2d_0^2{\bm\mu}^\dagger\cdot {\bf G}_{\rm b}({\bf r}_0,{\bf r}_0)\cdot{\bm\mu}/\hbar\epsilon_0$, 
where $d_0$ is the dipole moment of the optical transition. It is clear that at a C-point, a dipole aligned to the field for the forwards propagating Bloch mode, will be orthogonal to the field of the backwards propagating Bloch mode. Hence we find the following rate for spontaneous emission at a C-point:
\be
\Gamma_{\rm w}^{\rm C} =\Gamma_{\rm w}^{\rm f}=\frac{d_0^2 {\rm e}_0^2a\omega}{2 v_{g }\epsilon_0 \hbar}= \frac{d_0^2 \eta({\bf r}_0,\bm\mu)Q_{\rm w}}{\epsilon_0 \hbar V_{\rm eff} \epsilon_s},
\ee
where we have introduced an effective mode volume for the waveguide mode,
$V_{\rm eff} \equiv 1/(\epsilon_s |{\bf e}_{k}({\bf r}_0)|^2)$,
where the Bloch mode is at the antinode position,
and $\epsilon_s$ is the slab dielectric constant in which the QD is embedded. The waveguide mode decay rate is defined as $\kappa_{\rm w} =2v_g/a$,
so $Q_{\rm w}=\omega/\kappa_{\rm w}$. We have also introduced $\eta$; a spatial and polarization dependent function, varying between 0 and 1, to account for
deviations from the antinode and polarization coupling with the target PCW mode. In contrast, at a point where the polarisation is linear, and if the dipole is aligned to the field,  %
$\Gamma_{\rm w}^{\rm L} =\Gamma_{\rm w}^{\rm f}+\Gamma_{\rm w}^{\rm b}= 2\Gamma_{\rm w}^{\rm C}$. So despite the fact the dipole is aligned to the local field in both cases, the decay rate at the C-point is inherently half (assuming maximum coupling) of that at a point of linear polarisation. This is due to the lifting of the polarisation degeneracy between the forwards and backwards propagating modes, where at a C-point they are orthogonal. As such the density of available EM modes at a C-point is halved relative to a linear point where the local field contains no phase information. 
Using the PCW in Fig.\ref{fig:FDTD}, and assuming a realistic dipole moment of $d_0=30$ Debye we find a rate of emission for a spin-photon entangled source at a C-point of $\Gamma_{\rm w} \sim1.7$ GHz, corresponding to a Purcell factor of $P_f=\Gamma_{\rm w}/\Gamma^{\rm hom}=1.8$. This yields a beta factor of $\beta \sim 0.95$.

By allowing the spin to emit several photons in a row, large entangled photon states may easily be built up, useful for quantum metrology or one way quantum computation using the cluster state model \cite{Lindner:2009zr,nonlingate.pdf}.  The device may therefore operate as a pumped source (optically out-of-plane, or electrically) of entangled photons when the QD spin is located at the C-point. The C-point in a PCW is the only place in the waveguide where a QD spin may be used as a polarization/path entangled photon {\it source}, due to the perfect correlation of spin with path. Such device operation could never be predicted using a simple linear-dipole and LDOS approach commonly employed in cavity-QED.

As well as deterministic entangled photon sources, deterministic quantum gates would be a crucial component for scalable quantum devices. We now explore implications of considering polarization in PCWs when designing quantum circuits.  A PCW can be considered as a one dimensional waveguide as a result of lateral confinement by the photonic bandgap: photons are predominantly scattered either forwards and backwards in the waveguide itself. To perform a general analysis of the propagation and scattering of light in the PCW we again  take a Green function approach, where the  total field in the PCW, including the QD, and homogenous input field $\mathbf{E}^{\rm h}(\mathbf{r})$ may be expressed as
%
$\mathbf{E}(\mathbf{r})=\mathbf{E}^{\rm h}(\mathbf{r})
+ \mathbf{G}(\mathbf{r}, \mathbf{r}_{d})\cdot 
{\bm \alpha} \cdot {\bf E}^{\rm h}({\bf r}_0),
\label{eq:eprop2}
$ where 
${\bm \alpha} = \frac{\alpha_0\bm\mu\bm\mu^{\dag}}{1-\alpha_0 \bm\mu^{\dag} \cdot {\bf G}({\bf r}_0,{\bf r}_0) \cdot \bm\mu }$ is
is the QD polarizability, which includes 
coupling to the medium (while allowing for complex dipoles in a Cartesian coordinate system),  and the bare polarizability 
$
\alpha_0 = \frac{2 \omega_0 d_0^2/\epsilon_0\hbar}{\omega_0^2-\omega^2}
$, where we have neglected non-radiative losses.
%

Now consider a photon injected in the waveguide mode from the left (homogeneous solution),
$\mathbf{E}^{\rm h}(\mathbf{r})=\sqrt{\frac{a}{L}}
\mathbf{e}_{k_{\rm h}}(\mathbf{r})e^{i k_{\rm h} x }$. For a sufficiently long waveguide, the transmitted and reflected fields are given by 
$
\mathbf{E}_{\rm t}(\mathbf{r}; x\rightarrow \infty)=
\sqrt{\frac{a}{L}}\mathbf{e}_{k_{\rm h}}(\mathbf{r})e^{i k_{\rm h} x } 
+ \mathbf{G}_{\rm w}(\mathbf{r} ;x\rightarrow\infty, \mathbf{r}_{0})\cdot
{\bm \alpha} \cdot 
\sqrt{\frac{a}{L}}\mathbf{e}_{k_{\rm h}}(\mathbf{r}_0)e^{i k_{\rm h} x_{0} }$,
and 
$
\mathbf{E}_{\rm r}(\mathbf{r}, x\rightarrow -\infty)=
 \mathbf{G}_{\rm w}(\mathbf{r};x\rightarrow-\infty, \mathbf{r}_{d})\cdot
{\bm \alpha} \cdot 
\sqrt{\frac{a}{L}}\mathbf{e}_{k_{\rm h}}(\mathbf{r}_0)e^{i k_{\rm h} x_{0} },
$
%
where the only contribution from the total Green function far down the waveguide
is from the Bloch mode Green function (as we assume the QD is near the center of the waveguide). The transmitted and reflected amplitudes are, respectively,  given by $t (\omega)= {{\bf E}_{\rm t}(\mathbf{r}; x\rightarrow \infty)}/{{\bf E}^{\rm h}(\mathbf{r}; x\rightarrow \infty)}$ and   $r(\omega)= {{\bf E}_{\rm r}(\mathbf{r}; x\rightarrow -\infty)}/{{\bf E}^{\rm h}(\mathbf{r}; x\rightarrow -\infty)}$, which are derived to be
\be
t(\omega)= 1 + \frac{i\omega_02\Gamma_{\rm w}^{\rm f}}{\omega_0^2-\omega^2-i\omega_0
(\Gamma_{\rm w}^{\rm f}+\Gamma_{\rm w}^{\rm b}+\Gamma_0)},\\
\ee
and
\be
r(\omega) =  \frac{i\omega_02\Gamma_{\rm w}^{\rm f\rightarrow b} e^{2ik_{\rm h} x_0}}{\omega_0^2-\omega^2-i\omega_0
(\Gamma_{\rm w}^{\rm f}+\Gamma_{\rm w}^{\rm b}+\Gamma_0)},
\ee
where $\Gamma_{\rm w}^{\rm f\rightarrow b}$ is the scattering rate backwards given a forwards injected Bloch mode.  

\begin{figure}
\centering
\includegraphics[width=0.45\textwidth]{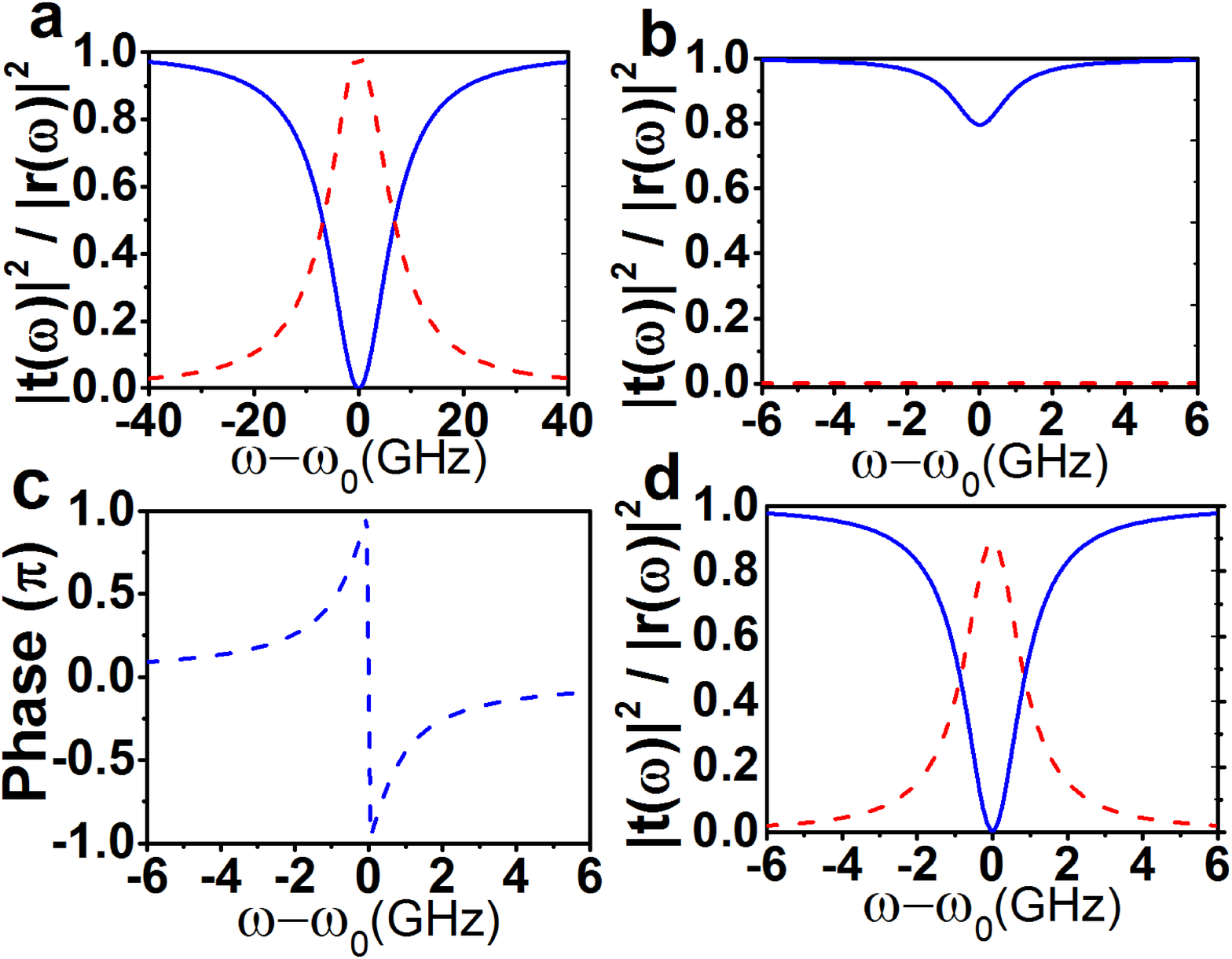}
\caption{(Color online) Transmitted (blue) and reflected (dashed red) intensity as a function of detuning for ({\bf a}) linear $E_y$ dipole placed at a point in the PCW with pure $E_y$ polarised light, ({\bf b}) $\sigma_+$ dipole at a $\sigma_+$ polarised C-point, with {\bf (c)} the accompanying  phase shift on the transmitted signal as a function of detuning. ({\bf d}) An $E_y$ dipole at a $\sigma_+$ polarised C-point. All plots use the W1 waveguide shown in Fig.~\ref{fig:FDTD}, with parameters $\Gamma_0=0.1\Gamma^{\rm hom}$ and $d_0=30$\,Debye.}\label{reflect}
\end{figure}

Now consider the case of a linearly polarised dipole, on an L-line in the PCW with the same linear polarisation (yellow line in Fig.~\ref{fig:FDTD}b). A photon with a narrow bandwidth relative to the dipole transition (weak excitation approximation) input into the forwards propagating waveguide mode leads to the frequency dependent response in Fig.~\ref{reflect}a. On resonance ($\omega=\omega_0$), the dipole will scatter with the rates $\Gamma_{\rm w}^{\rm f}=\Gamma_{\rm w}^{\rm b}=\Gamma_{\rm w}^{\rm f\rightarrow b}$. Hence $|t(\omega)|^2\approx0$, and $|r(\omega)|^2\approx1$, and scattering from a QD leads to reflection back along the waveguide as predicted in earlier works~\cite{Shen:2005nx}. One observes a dipole-induced-reflection \cite{Kochan:1994uq} identical to that in a cavity-waveguide architecture \cite{dipoleRef,waks:153601}. The dipole induced reflection feature in Fig.~\ref{reflect}a. has a width of $\sim14\,$GHz based on the waveguide simulated in Fig.~\ref{fig:FDTD} again assuming a $d_0=30$~Debye.  This compares favourably with drop filter cavity designs~\cite{waks:153601}, where the transparency window has a width of $\sim100$ GHz. Optimisations away from the standard W1 waveguide should result in the transparency window becoming even wider. Again if we consider a charged QD; by initialising in the spin up state $\ket{\!\uparrow}$, a resonant photon injected into the forwards propagating mode after scattering will end up in the entangled state:
\be
\ket{\psi}=\ket{b}\ket{+}+\ket{f}\ket{-}
\ee
\noindent where $\ket{+} =\ket{\!\uparrow}+\ket{\!\downarrow}$, and $\ket{-} =\ket{\!\uparrow}-\ket{\!\downarrow}$ represent the spin in the computational basis. By performing single qubit rotations on the spin one can arrive at the same entangled state in Eq.~(\ref{cpem}) for a charged QD emitting light at a C-point. Also, since along L-lines the local field has no fixed phase relation between $E_x$ and $E_y$, the local field at the QD location (${\bf r}_0$) is the same in both forwards and backwards propagating directions, i.e.,  ${\bf e}_k({\bf r}_0)={\bf e}^*_k({\bf r}_0)$. This allows one to encode photons via their path ($\ket{f}$ or $\ket{b}$) and realise a fully deterministic spin photon interface~\cite{PhysRevLett.92.127902, PhysRevLett.104.160503, PhysRevB.78.085307}.

If we now move to a point where the local polarisation is circular then one sees a significant departure from the above. Figure \ref{reflect}b is a plot of the behaviour for a right circularly polarised dipole at a C-point (yellow circle in Fig.~\ref{fig:FDTD}b). We again look at the output response to a photon input into the forwards propagating mode as a function of detuning from the dipole frequency. Since we inject photons into the forwards propagating mode the field created at the dipole location (${\bf r}_0$) is $\sigma_+$ polarised. For the case when the dipole is also $\sigma_+$ polarised then we find that $\Gamma_{\rm w}^{\rm b}=\Gamma_{\rm w}^{\rm f\rightarrow b}=0$, with on resonance excitation and  $\Gamma_0=0.1\Gamma^{\rm hom}$, then $|r(\omega)|^2\approx0$ and $|t(\omega)|^2\approx0.8$. In this instance no light is reflected but is transmitted with a $\pi$ phase shift due to the interaction with the dipole. The reduction in the transmitted intensity is due to out of plane scattering. At the C-point considered here, we find $\eta({\bf r}_0,\bm\mu)\sim0.25$ as the C-point is not at a field antinode. Optimising the PCW structure to increase $\eta({\bf r}_0,\bm\mu)$ will increase $\Gamma_{\rm w}^{\rm f}$, improving the $\beta$-factor to give near unit transmission with a $\pi$ phase shift. If the dipole is   $\sigma_-$ polarised, then $\Gamma_{\rm w}^{\rm f}=\Gamma_{\rm w}^{\rm f\rightarrow b}=0$, i.e., there is no interaction and the photon transmits without a phase shift.
Considering a simple two level system model, if the dipole is in an equal superposition of $\sigma_+$ and $\sigma_-$ (linear), then we predict $\Gamma_{\rm w}^{\rm f}=\Gamma_{\rm w}^{\rm b}=\Gamma_{\rm w}^{\rm f\rightarrow b}$, and at the dipole resonance $|t(\omega)|^2\approx0$, $|r(\omega)|^2\approx0.9$ as in Fig.~\ref{reflect}c. Now we find that we see a zero in transmission and a reflection as a result of scattering from the dipole. This is caused by destructive interference between the $\sigma_+$ and $\sigma_-$ components in the forwards propagating direction. This is exactly the same as in Fig.~\ref{reflect}a except the bandwidth and intensity of the dipole induced reflection feature is reduced. This is due to polarisation mismatch and because the C-point is moved from the antinode of the Bloch mode, giving $\eta({\bf r}_0,\nr)\sim0.125$. 

Again considering the behaviour of a charged QD at the C-point, if the spin is $\ket{\!\downarrow}$, corresponding to a $\sigma_-$ polarised dipole, then as above there is no interaction and a forwards injected resonant photon will transmit. If the spin is $\ket{\!\uparrow}$, i.e. a $\sigma_+$ dipole transition, then the light transmits with a $\pi$ phase shift. If we prepare the QD spin in an equal superposition of $\sigma_+$, and $\sigma_-$ (i.e., $\ket{\uparrow}+\ket{\downarrow}$), then after interaction with a forwards injected resonant photon we have the state,
\be
\ket{\psi}_{\rm out}=-\ket{f}\ket{\!\uparrow}+\ket{f}\ket{\!\downarrow}.
\ee
where we have now set $\Gamma_0=0$ for simplicity. This output state clearly does not correspond with the semiclassical result for a simple two level system in Fig.~\ref{reflect}c, since there is no longer an available backwards propagating photon state. It is clear from this equation that the addition of spin into the system prevents destructive interference in the forwards propagating direction. The charged QD system can never give rise to a reflection at a C-point. This is in contrast to a fine structure split neutral QD where there is no ground state spin and the linear transitions would give rise to a reflection at a C-point. Further if we were to input incoherent photons into the forwards propagating mode and set $\ket{\psi}_{\rm spin}=\ket{\uparrow}+\ket{\downarrow}$, then one would detect output photons in the forwards and backwards mode with equal probability. This result highlights the role that coherence and quantum entanglement can play in light matter interactions where in this example, surprisingly, it completely reverses the direction of light propagation. 

In conclusion we have demonstrated, using a rigorous  Green function method, that the projected LDOS in complex nanophotonic structures such as PCWs has important phase information that must not be neglected.  We demonstrate the importance of this by considering a QD spin emitter in a PCW, and show that one may control the direction of photon emission by controlling the spin orientation. Entangled photon sources may be generated at a C-point polarization singularity whilst at both C-points and L-lines one may entangle photons via dipole induced reflection, all with $>90\%$ efficiency. Most importantly, we develop  a general and intuitive mathematical framework to understand the interaction between dipoles and fields in chiral photonic structures, and show the limitations of a semiclassical analysis, where quantum entanglement can completely reverse the photon propagation direction.

\acknowledgements

The authors acknowledge helpful discussions with P. Lodahl, B. Lang, and R. Ge. This work was carried out using the computational facilities of the Advanced Computing
Research Centre, University of Bristol Ð http://www.bris.ac.uk/acrc/. This work has been
funded by the project ÒSPANGL4QÓ, under FET-Open grant number: FP7-284743. RO was
sponsored by the EPSRC under grant no. EP/G004366/1, and JGR is sponsored under ERC Grant No. 247462 QUOWSS. This work  is part of the research program of the Stichting voor Fundamenteel Onderzoek der Materie, which is financially supported by the Nederlandse Organisatie voor Wetenschappelijk Onderzoek. DMB acknowledges support from a Marie Curie individual fellowship, and SH acknowledges funding from the Natural Sciences and Engineering Research Council of Canada. 

{\it Note added.} After submission  we became aware of two related works: Ref.~\onlinecite{sollner:arxiv} considers a CNOT gate implementation in similar structures, and Ref.~\onlinecite{Mitsch:arxiv} shows directionality of emission from single atoms coupled to optical fiber.

\end{document}